\documentstyle[12pt]{article}
\begin{document}
\def\la{\mathrel{\mathchoice {\vcenter{\offinterlineskip\halign{\hfil
$\displaystyle##$\hfil\cr<\cr\sim\cr}}}
{\vcenter{\offinterlineskip\halign{\hfil$\textstyle##$\hfil\cr
<\cr\sim\cr}}}
{\vcenter{\offinterlineskip\halign{\hfil$\scriptstyle##$\hfil\cr
<\cr\sim\cr}}}
{\vcenter{\offinterlineskip\halign{\hfil$\scriptscriptstyle##$\hfil\cr
<\cr\sim\cr}}}}}
\title{\large \hspace{10cm} ITEP-31/98 \\ \hspace{10cm} August 1998 \\
\vspace{1cm}
\LARGE \bf
Kink-antikink interactions in the double sine-Gordon equation and
the problem of resonance frequencies}
\author {V. A. GANI\thanks{E-mail: gani@vitep5.itep.ru}{\,}
\\
{\it Moscow State Engineering Physics Institute (Technical University),}\\
{\it Kashirskoe shosse, 31, Moscow, 115409, Russia}\\
{\it and}\\
{\it Institute of Theoretical and Experimental Physics, Russia}\\
\\
A. E. KUDRYAVTSEV \thanks{E-mail: kudryavtsev@vitep5.itep.ru}
\\
{\it Institute of Theoretical and Experimental Physics,}\\
{\it B.Cheremushkinskaja, 25, Moscow, 117259, Russia}\\
}
\date{}
\maketitle
\vspace{1mm}
\centerline{\bf {Abstract}}
\vspace{3mm}

     We studied the kink-antikink collision process for the "double
sine-Gordon" (DSG) equation in 1+1 dimensions at different values of the
potential parameter $R>0$. For small values of $R$ we discuss the problem
of resonance frequencies. We give qualitative explanation of the frequency
shift in comparison with the frequency of the discrete level in the
potential well of isolated kink. We show that in this region of the
parameter $R$ the effective long-range interaction between kink and
antikink takes place.

\newpage

\begin{center}
\bf
1. Introduction
\end{center}

     Resonant energy exchange mechanism, that we shall consider in this our
paper, was originally observed in the kink-antikink collisions for the
$\lambda\phi^4_2$-theory. To examine such process one should consider
initial configuration in the form of kink ($K$) and antikink ($\bar{K}$)
placed at $x=\pm x_0$ ($x_0\gg1$) moving toward each other with some
velocities $v_i$. It was found that there is critical value of the initial
velocity $v_{cr}\cong0.2598$ and at $v_i>v_{cr}$ inelastic $K\bar{K}$
scattering takes place while at $v_i<v_{cr}$ kink and antikink form a bound
state. This bound state then decays into small oscillations~[1].

     Later on, when the $K\bar{K}$ collision process was studied more
careful, so-called escape windows were found~[2] in the range of the initial
velocities $v_i<v_{cr}$. Escape windows are nothing more than some values
of the initial velocity $v_i=v_n$ at which kinks escape to infinity after
second collision instead of forming a bound state. This phenomenon was
semiquantitative
explained in Ref.~[2]. The point is that the $\lambda\phi^4_2$-theory kink
excitation spectrum has one zero (translational) and one non-zero (shape)
mode with the frequency $\omega_1=\sqrt{3/2}$. It was observed that the
following condition is satisfied with a reasonable accuracy:
$$
\omega_1T_{12}(v_n)=\delta+2\pi n \ , \eqno(1)
$$
where $T_{12}$ is the time interval between the two collisions of the kinks,
$n$ is integer, $\delta$ is some constant phase.
During the first $K\bar{K}$ collision a part of their kinetic energy is
transferred to excitation of the kink discrete mode $\omega_1$. Therefore,
kinks cannot escape to infinity and only go away at some distance and
collide again. If condition (1) is satisfied, part of energy which conserved
in the mode $\omega_1$, is returned back to kinks translational mode
(kinetic energy) and kinks can overcome the mutual attraction and go to
infinity. Just this phenomenon was named "resonant energy transfer mechanism".

     Note, that "higher orders" escape windows were also found. In these
cases $K$ and $\bar{K}$ escape to infinity after three or more collisions.
For more detailed information about solitary wave interactions in the
classical field theory see Review~[3].

     Let's now turn our attention to the system which we shall investigate
in the present paper. The double sine-Gordon (DSG) equation can be obtained
from the Lagrangian of the form
$$
{\cal L}=\frac{1}{2}\left(\frac{\partial\phi}{\partial t}\right)^2-
\frac{1}{2}\left(\frac{\partial\phi}{\partial x}\right)^2-V(\phi)
\eqno(2)
$$
with the potential
$$
V(\phi)=-\frac{4}{1+4|\eta|}\left(\eta\cos\phi-\cos\frac{\phi}{2}\right) \ .
\eqno(3)
$$
Parameter $\eta$ may be assigned any arbitrary real value
($-\infty<\eta<+\infty$). From Lagrangian (2) we get for the real scalar
field $\phi(x,t)$ in (1+1) dimensions the following equation:
$$
\frac{\partial^2\phi}{\partial t^2}-\frac{\partial^2\phi}{\partial x^2}+
\frac{2}{1+4|\eta|}\left(2\eta\sin\phi-\sin\frac{\phi}{2}\right)=0 \ .
\eqno(4)
$$
In the present work we shall consider the range $\eta>0$. In this case it is
suitable to introduce parameter $R$ related with $\eta$ by the equality:
$$
\eta=\frac{1}{4}\sinh^2R \ .
$$
Equation (4) has static topological solution in the form of $4\pi$-kink
(antikink):
$$
\phi_{K(\bar{K})}(x)=4\pi n\pm4\arctan\frac{\sinh x}{\cosh R} \ .
\eqno(5)
$$
The sign "$+$" corresponds to the case of kink, "$-$" -- to the case of
antikink, $n$ is integer.
Eq.~(5) can be rewritten in the form:
$$
\phi_{K(\bar{K})}(x)=4\pi n\pm[\phi_{SGK}(x+R)-\phi_{SGK}(R-x)] \ ,
\eqno(5a)
$$
where $\phi_{SGK}(x)=4\arctan\exp(x)$ is the sine-Gordon (SG) equation
$2\pi$-soliton. From Eq.~(5a) the physical meaning of the parameter $R$
becomes clear: DSG kink can be interpreted as a superposition of two SG
solitons, separated by the distance $2R$.

     The $K\bar{K}$ collision process at a variety of values of the parameter
$\eta$ and the initial velocity $v_i$ was studied in details in Ref.~[4].
As for the $\lambda\phi^4_2$-theory case, there is some critical velocity
$v_{cr}$ below which kinks form a bound state decaying into small
oscillations. Note, that in the DSG case this critical velocity is
a function of the parameter ($\eta$ or $R$)~[4].

     It was found that in the DSG system the resonant energy exchange
mechanism also takes place.
As a consequence there is a system of escape windows at some values
of $\eta$. Note, that there is one important difference in kinks collision
processes between
$\lambda\phi^4_2$ and DSG models. In the first case kinks cannot pass
through each other at a valuable distance, while in the second they can
travel to infinity after passing through each other. This difference is
a consequence of the different structure of the potential $V(\phi)$.

     In Ref.~[4] different values of the parameter $R$ were studied.
At $R=1.2$ a typical picture of escape windows was similar to the
$\lambda\phi^4_2$-theory case. However, at smaller $R$, namely at $R=0.5$,
a new phenomenon was observed in the $K\bar{K}$ collisions -- so-called
quasiresonances.
The essence of the phenomenon is in the following. At all velocities
$v_i<v_{cr}$ we get capture and formation of $K\bar{K}$ bound state.
But the time between the second and third collisions $T_{23}$ as a function
of the initial velocity $v_i$ has a series of well-defined maxima,
see Fig.~1. Such behavior of $T_{23}(v_i)$ means that the resonant energy
exchange mechanism appears in the system, but at the same time the energy
returned to the translational mode during the second collision is not enough
for kinks to escape to infinity after the second collision. Besides, it turned
out that the frequency of oscillations in which a part of kinetic energy
is transferred while kinks passing through each other for the first time,
i.e. $\omega_1$ in Eq.~(1), is smaller than the frequency of the
localized DSG kink excitations at given $R$.

\begin{center}
\bf
2. General approach
\end{center}

     Up to now, investigating the resonant energy exchange mechanism, people
involved the localized excitations over an isolated kink (or antikink).
As it will be shown, in some cases of the $K\bar{K}$ scattering such
approximation is not valid. In these cases one should consider the spectrum
of small excitations for the $K\bar{K}$ system as a whole. For this purpose
let's look for the solution of Eq.~(4) for the field $\phi$ in the form:
$$
\phi(x,t)=2\pi+\phi_K(x-x_0)+\phi_{\bar{K}}(x+x_0)+\delta\phi(x,t) \ .
$$
Such configuration corresponds to the kink and antikink placed at
$x=\pm x_0$ plus some small perturbation $\delta\phi(x,t)$, $|\delta\phi|\ll1$.
Taking into account that $\phi_K$ and $\phi_{\bar{K}}$ are solutions of
Eq.~(4), we get for $\delta\phi$ the following linearized equation:
$$
\delta\phi_{tt}-\delta\phi_{xx}+\delta\phi\cdot\left.
\frac{\partial^2V}{\partial\phi^2}
\right|_{\phi=2\pi+\phi_K+\phi_{\bar{K}}}=Q(x,x_0) \ ,
\eqno(6)
$$
where
$$
Q(x,x_0)=
\left.\frac{\partial V}{\partial\phi}
\right|_{\phi=\phi_K}+
\left.\frac{\partial V}{\partial\phi}
\right|_{\phi=\phi_{\bar{K}}}-
\left.\frac{\partial V}{\partial\phi}
\right|_{\phi=2\pi+\phi_K+\phi_{\bar{K}}} \ .
\eqno(7)
$$
The explicit form of the function $Q(x,x_0)$ is rather cumbersome
(see Appendix),
but nevertheless we can make several general notes. Inhomogeneity $Q(x,x_0)$
in Eq.~(6) is a consequence of the fact that the configuration "kink+antikink"
is not a solution of Eq.~(4). The function $Q(x,x_0)$ characterizes
overlapping of the kink and antikink, because $\phi_K(x-x_0)$ and
$\phi_{\bar{K}}(x+x_0)$ are exact solutions of Eq.~(4) when taken separately.
Obviously, $Q(x,x_0)$ is an even function of $x$ and $x_0$ and it falls down
exponentially when $x_0$ increases. At fixed $x_0$ as a function of $x$ \
$Q(x,x_0)$ looks like two bumps with maxima at $x=\pm x_0$.

     Let's now find the excitation spectrum for $\delta\phi$. For this
purpose we take Eq.~(6) with zero right-hand side and look for $\delta\phi$
in the form:
$$
\delta\phi(x,t)=e^{i\omega t}\chi(x) \ .
$$
Then for the function $\chi(x)$ we get the following differential equation
of the Schr\"odinger type:
$$
-\chi^{\prime\prime}+U(x,x_0)\chi=\omega^2\chi \ ,
\eqno(8)
$$
where
$$
U(x,x_0)\equiv\left.\frac{\partial^2V}{\partial\phi^2}
\right|_{\phi=2\pi+\phi_K+\phi_{\bar{K}}} \ .
\eqno(9)
$$
The explicit form of the potential $U(x,x_0)$ is rather complicated
(see Appendix) and depends crucially on $x_0$. Note, that the shape of this
potential depends on the parameter ($R$ or $\eta$) and $U(x,x_0)\to1$ when
$x\to\pm\infty$. Hence, $\omega<1$ form the discrete excitation spectrum,
and $\omega>1$ -- the continuum one. In the limit
$x_0\gg1$ \ $U(x,x_0)$ as a function of $x$ looks like two identical potential
wells, separated by the distance $2x_0$. Each well contains one or more
discrete levels which correspond to the localized excitations of the solitary
kink (antikink).
In the collision process DSG kinks pass through each other, i.e. $x_0$
decreases to zero and then starts to increase again. At small $x_0$ the
distance between the wells is small and the discrete levels are not
independent. With kinks moving toward each other from the infinity the levels
begin to split and then at $x_0\la1$ mutual potential of the system
$K\bar{K}$ is quite different from one of the solitary kink (antikink).

     It is worth mentioning that taking into account of both wells is
also necessary in cases when in each potential well there is a discrete
level with small binding energy situated near the continuum.
In such cases one should take into account overlapping of the wave functions
in both wells even at $x_0\gg1$. It means that under some conditions
long-range interaction between kink and antikink appears in the system.

     In what follows we will show, that within such approach the phenomenon
of quasiresonances observed in the DSG system at $R=0.5$ in
Ref.~[4] may be simply explained. We will also argue that the cause of
the quasiresonances is just the resonant energy exchange mechanism, that
leads to escape windows at some other values of $R$. Moreover, there is
some intermediate region of $R$ where quasiresonances and escape windows
appear together.

\begin{center}
\bf
3. Small $R$
\end{center}

     In Ref.~[4] quasiresonances were observed at $R=0.5$. We performed
similar calculations and obtained analogous curve $T_{23}(v_i)$, see Fig.~1.
Besides that, we have investigated the $K\bar{K}$ collision process at
$R=0.4$ and $R=0.6$. At $R=0.4$ (Fig.~2) we get a picture of quasiresonance
peaks analogous to the case $R=0.5$. At $R=0.6$ (Fig.~3) there seem to exist
escape windows in places of some peaks on the curve $T_{23}(v_i)$. It confirms
that quasiresonances and escape windows are phenomena of the same nature and
with the parameter $R$ increasing some quasiresonance peaks transform into
the escape windows. At some intermediate values of $R$ both phenomena are
presented, and in further increasing of $R$ only escape windows survive.
At $R=1.2$ in Ref.~[4] a perfect picture of escape windows and no
quasiresonances were observed.

     To answer the question why at given $R$ quasiresonances or escape
windows appear it is required, generally speaking, to solve Eq.~(6) for
$\delta\phi$ with the right-hand side. At the same time we can suggest
some true-like hypothesis.
Each bump of the source $Q(x,x_0)$ is localized on size of
order of 1 (see Appendix). In the case of small $R\sim0.4-0.6$ the first
excited level in the well is not well-localized (binding energy is small).
Therefore, integral of overlapping of $Q(x,x_0)$ and the wave function of
the excited state is small. It corresponds to the fact that the part
of the kinetic energy transferred to the discrete mode $\omega_1$ is small,
and hence loss of energy due to radiation is large. The situation changes
with increasing of $R$. The binding energy of the first excited level is
increasing, and for $R=1.2$ the first excited level in the well is already
well-localized. Because of this reason the character size of the wave
function is of the same order as the source one. Hence the energy transfer
mechanism is more effective in this case.

     From analysis of the quasiresonance peaks of the $T_{23}(v_i)$ plot
for $R=0.5$ (Fig.~1) it follows that the frequency of the discrete mode being
excited is approximately equal to $\tilde{\omega}_1=0.945$. At the same time
in the well corresponding to one kink excitations there is discrete level with
frequency $\omega_1=0.967$.
As it will be shown, this deviation is not incidental and may be easily
interpreted within our approach.

     At small $R$ the system is close to the pure sine-Gordon. Therefore,
the critical velocity $v_{cr}$ is small ($v_{cr}=0$ corresponds to the
pure sine-Gordon) and potential (9) in the Schrodinger equation (8) has
one discrete level situated near the continuum (in the pure sine-Gordon
case there is only zero mode). Presence of a shallow level implies that the
corresponding wave function falls down slowly with the distance from the well.
Due to this reason while studying the $K\bar{K}$ collision process it is
necessary to take into account the fact that the wells affect each other
even at large distance. It leads to changes in the excitation spectrum.
In Fig.~4 we show how the excitation frequency of the $K\bar{K}$ system
depends on the distance between $K$ and $\bar{K}$ (this distance is equal
to $2x_0$). From the plot it is seen that even at $x_0\gg1$ there present
some visible splitting of the higher discrete level.
In presence of the second well this
level $\omega_1=0.967$ splits into two sublevels: the higher with
$\tilde{\omega}_1^{odd}>0.967$ and the lower with
$\tilde{\omega}_1^{even}<0.967$. In the collision process
the lower one is excited because the corresponding wave function is even.
Moreover, the higher level may disappear during the collision. From Fig.~4
one can see, that this higher level appears from the continuum
$\omega>1$ at some critical distance $2(x_0)_{cr}\gg1$ between kink and
antikink. So, if the original level in a single well lies near the continuum
(as it happens at small $R$), then $(x_0)_{cr}$ is very large.
At the same time the level with $\tilde{\omega}_1^{even}$ exists in
a relatively wide interval of distances $x_0$ between kink and antikink.
Namely this level is excited during the $K\bar{K}$ collision because of
the resonant energy exchange mechanism. In this case the frequency
$\omega_1$ in expression (1) is indeed the averaged over different $x_0$
the frequency $\tilde{\omega}_1^{even}$. It is smaller than the frequency
of the discrete mode of isolated kink, what is in correspondence with the
numerical simulations.

\begin{center}
\bf
4. Numerical calculations
\end{center}

     We solved the second order partial differential equation (4)
numerically on the lattice with $\Delta x=0.01$. Initial conditions were
taken in the form of kink and antikink (5) situated at $x=\pm20$
moving towards each other with velocities $\mp v_i$ respectively. Moments
of kink and antikink passing through each other were fixed via field behavior
at the origin $x=0$.

     To find discrete levels in the potential (9) we used the fact that
the wave function falls down exponentially at large distances. We took
solution
of the Schrodinger equation in the form $\chi\sim\exp{(x\sqrt{1-\omega^2})}$
at $x=-50$ and solved numerically stationary equation (8). As a result we got
$\chi$ at $x=50$ as a function of $\omega$. When $\omega$ does not correspond
to the discrete level, $\chi$ grows exponentially with $x$ at positive $x$'s.
But if $\omega$
coincides with a discrete level of the potential, then $\chi$ is exponentially
suppressed at large $x$. In real computations we observed that $\chi(x=50)$
changed its sign when $\omega$ passed a discrete level. Note, that this method
being applied to searching of a shallow level does not return a good result.
In such a case one should take more distant starting and ending points.
The origin of the problem is in the following: for a shallow level
$\omega\to1$, and exponents fall and grow very slowly with increasing of $x$.

\begin{center}
\bf
Conclusion
\end{center}

     This our paper presents qualitative and semiquantitative explanation
of the phenomenon of quasiresonances in collisions of kink and antikink
of the double sine-Gordon equation at small $R$. It is shown that the
resonant energy exchange mechanism being applied in its previous form
gives not satisfactory results for frequencies.

     It was shown that the resonant energy exchange between kinks'
translational mode and the discrete excitations of the $K\bar{K}$
system as a whole takes place. At small $R$ it is essential because of
long-range interaction in the system caused by the presence of a shallow level
in the discrete spectrum of excitations of an isolated kink (antikink).

     The proposed mechanism explains qualitatively the decrease of the
resonance frequency $\omega_1$ in Eq.~(1) at small $R$ in comparison
with the discrete frequency of an isolated kink.

\begin{center}
\bf
Acknowledgments
\end{center}

     We are thankful to Dr. V.~G.~Ksenzov for his interest to our work and to
Professor W.~J.~Zakrzewski for helpful discussion. One of the authors
(V.~A.~Gani) is indebted to A.~A.~Panfilov for discussing several questions
related to the numerical calculations.

     This work was supported in part by the Russian Foundation for Basic
Research under Grant No~98-02-17316 (both authors) and under Grant
No~96-15-96578 (A.~E.~Kudryavtsev). The work of V.~A.~Gani was also supported
by the INTAS Grant No~96-0457 within the research program of the International
Center for Fundamental Physics in Moscow. One of the authors
(A.~E.~Kudryavtsev) would also like to thank RS Grant for finansial support.

\newpage

\begin{center}
\bf
Appendix
\end{center}

     In equation
$$
\frac{\partial^2\phi}{\partial t^2}-\frac{\partial^2\phi}{\partial x^2}+
\frac{\partial V}{\partial\phi}=0 \ ,
\eqno(A1)
$$
where $V(\phi)$ has the form (3) let us substitute
$\phi=2\pi+\phi_K+\phi_{\bar{K}}+\delta\phi$.
Taking into account that $\phi_K$ and $\phi_{\bar{K}}$ are exact solutions of
(A1) and linearizing with respect to $\delta\phi$ we get:
$$
\frac{\partial^2\delta\phi}{\partial t^2}-
\frac{\partial^2\delta\phi}{\partial x^2}+
\left.\frac{\partial^2V}{\partial\phi^2}\right|_{2\pi+\phi_K+\phi_{\bar{K}}}
\delta\phi=Q(x,x_0) \ ,
\eqno(A2)
$$
where $Q(x,x_0)$ is given by Eq.~(7).
If we substitute explicit expressions for kink and antikink situated at
$x=\pm x_0$ respectively, then we get:
$$
Q(x,x_0)=
\frac{8}{1+4\eta}
\left[\frac{s_+-s_-}{(1+s_+^2)(1+s_-^2)}
\right.
$$
$$
\left.
+4\eta\left[
\frac{s_-(1-s_-^2)}{(1+s_-^2)^2}\left(1-\left(\frac{1-s_+^2}{1+s_+^2}\right)^2
\right)-
\frac{s_+(1-s_+^2)}{(1+s_+^2)^2}\left(1-\left(\frac{1-s_-^2}{1+s_-^2}\right)^2
\right)\right]\right]
\eqno(A3)
$$
here
$$
s_{\pm}=\frac{\sinh{(x\pm x_0)}}{\cosh{R}}
$$
($\eta$ and $R$ are related by $\eta=(1/4)\sinh^2R$.)

     Let us use $Q(x,x_0)=0$ in (A2) and substitute
$\delta\phi(x,t)=\chi(x)\exp{(i\omega t)}$, then we obtain:
$$
-\chi^{\prime\prime}+U(x,x_0)\chi=\omega^2\chi \ ,
$$
where
$$
U(x,x_0)\equiv\left.\frac{\partial^2V}{\partial\phi^2}
\right|_{\phi=2\pi+\phi_K+\phi_{\bar{K}}} \ .
$$
Insert here explicit expressions for kink and antikink. As a result we have
$$
U(x,x_0)=\frac{1}{1+4\eta}
\left[
\frac{1-s_-^2}{1+s_-^2}\frac{1-s_+^2}{1+s_+^2}
+\frac{4s_-s_+}{(1+s_-^2)(1+s_+^2)}
\right.
$$
$$
\left.
+8\eta\left[
\frac{1-s_-^2}{1+s_-^2}\frac{1-s_+^2}{1+s_+^2}
+\frac{4s_-s_+}{(1+s_-^2)(1+s_+^2)}
\right]^2-4\eta\right] \ .
\eqno(A4)
$$

\newpage

\begin{center}
\bf
Figure captions
\end{center}
\bigskip

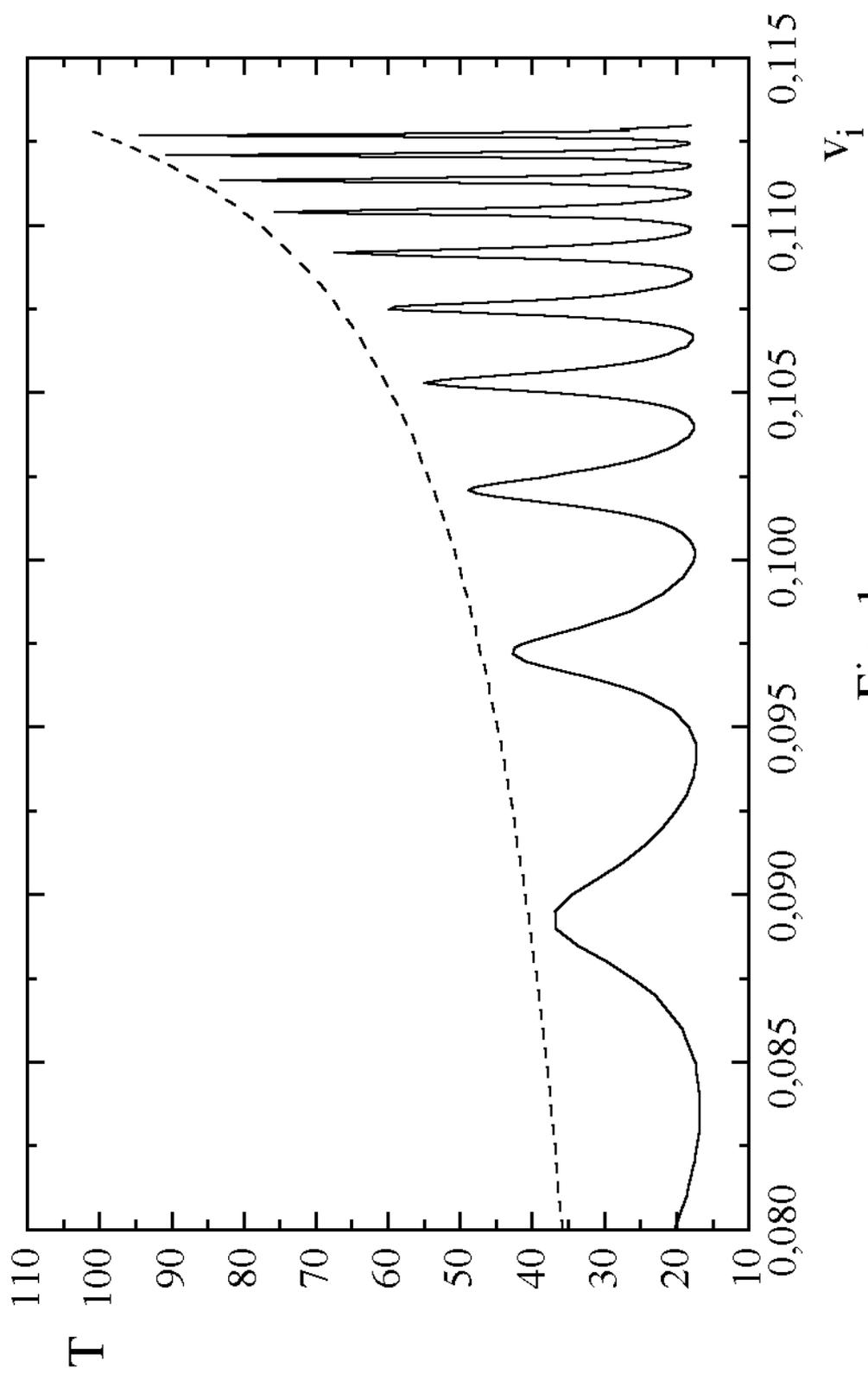
{\bf Fig.~1.} The time $T_{12}$ between the first two $K\bar{K}$ collisions
        (dashed curve) and the time $T_{23}$ between their second and
        third collisions (solid curve) as functions of the initial
        velocity $v_i$ for $R=0.5$.

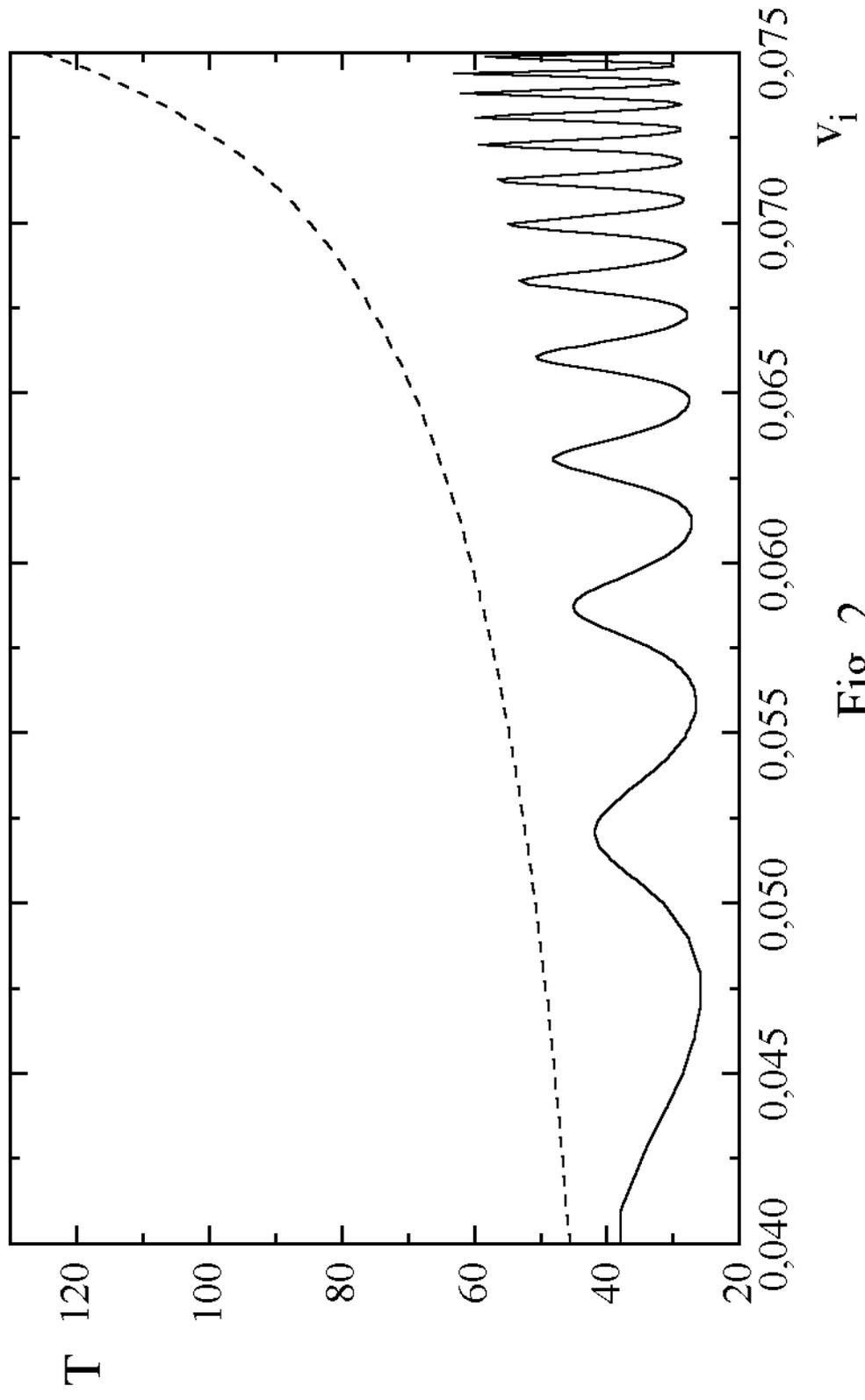
{\bf Fig.~2.} The time $T_{12}$ between the first two $K\bar{K}$ collisions
        (dashed curve) and the time $T_{23}$ between their second and
        third collisions (solid curve) as functions of the initial
        velocity $v_i$ for $R=0.4$.

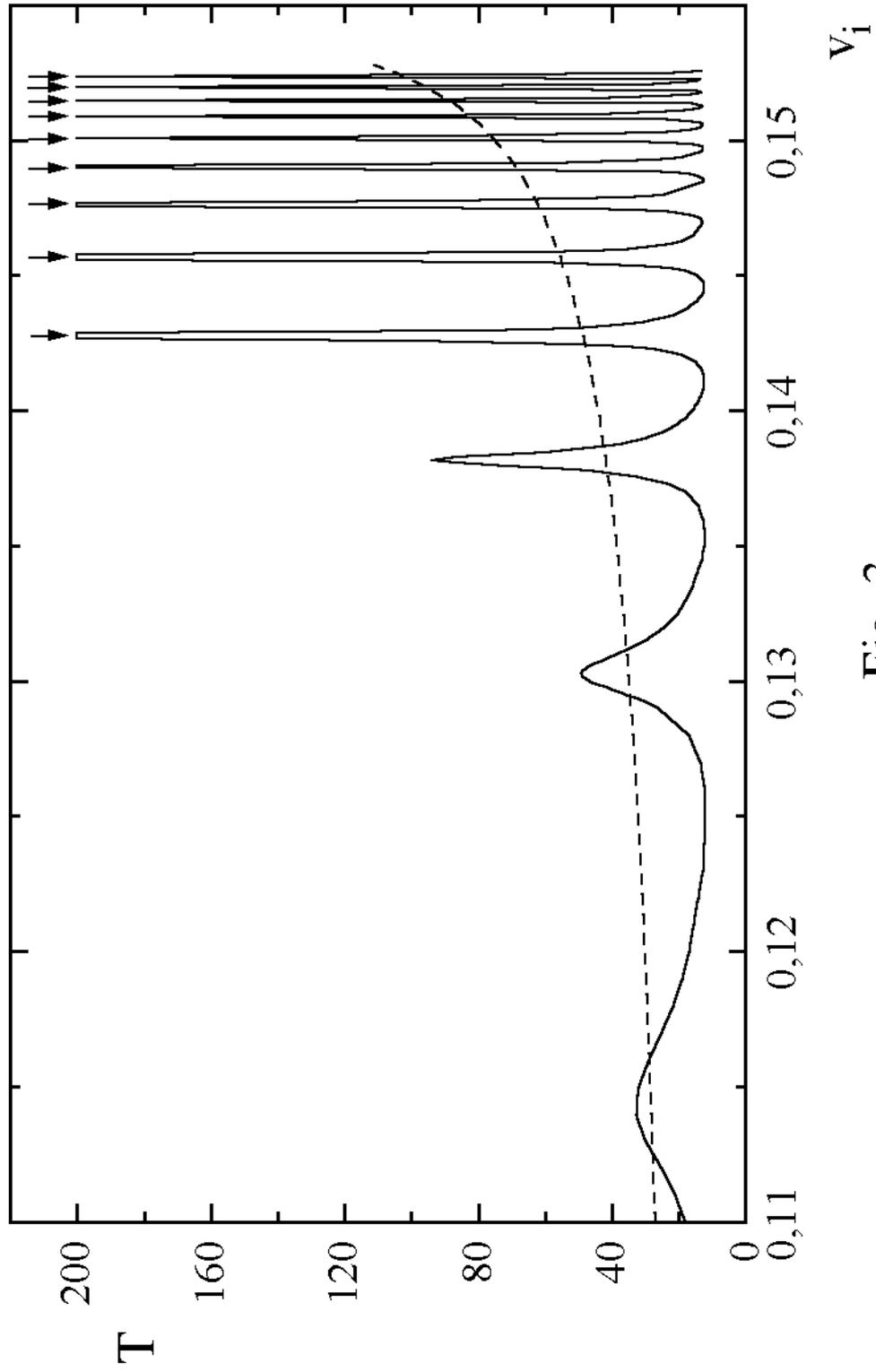
{\bf Fig.~3.} The time $T_{12}$ between the first two $K\bar{K}$ collisions
        (dashed curve) and the time $T_{23}$ between their second and
        third collisions (solid curve) as functions of the initial
        velocity $v_i$ for $R=0.6$. Arrows denote probable positions of
        the escape windows.

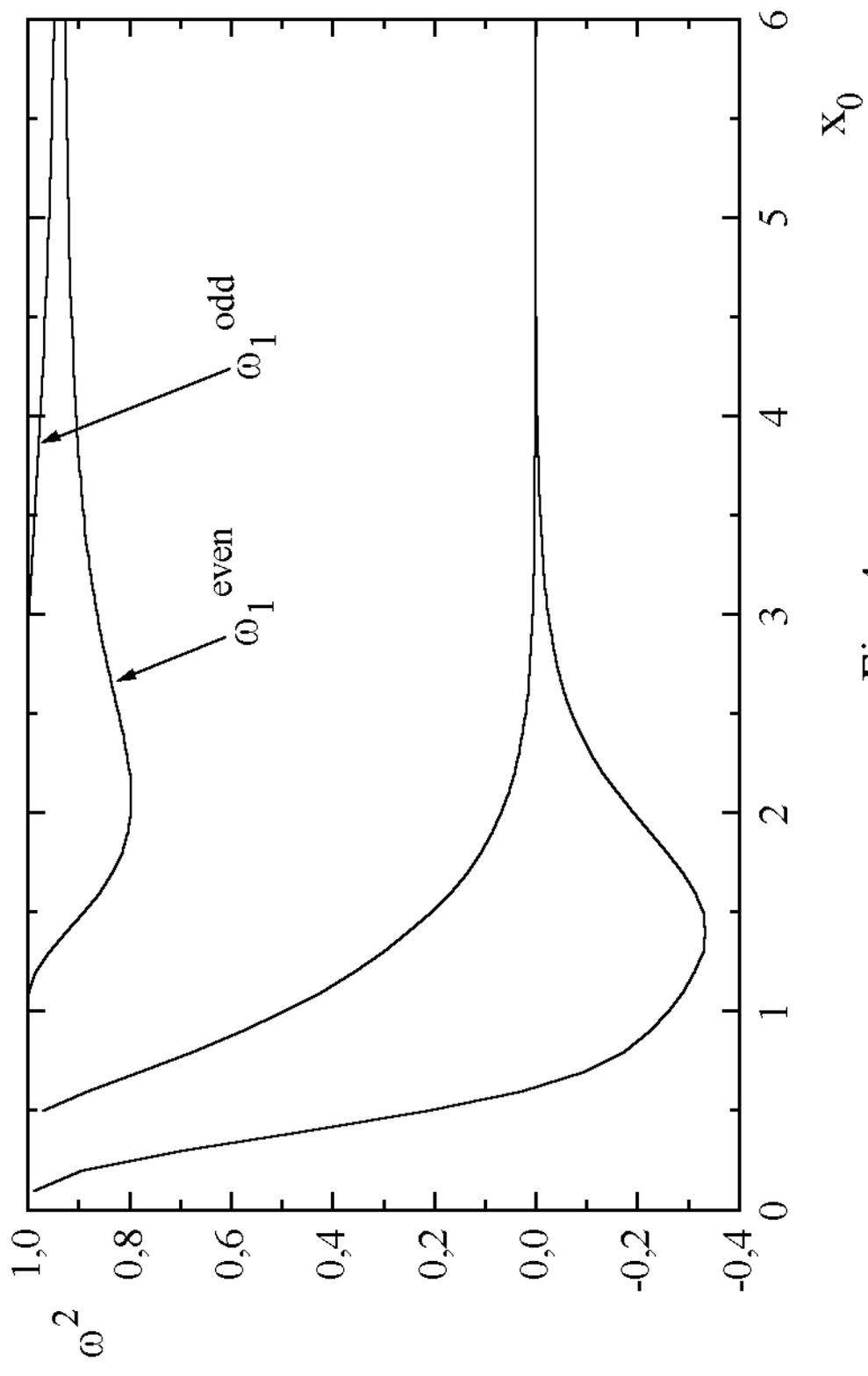
{\bf Fig.~4.} The excitation frequency as a function of the initial
              half-distance between $K$ and $\bar{K}$. $R=0.5$.


\begin{thebibliography}{100}
\bibitem{1} A.~E.~Kudryavtsev, Pis'ma Zh. Exp. Teor. Fiz., {\bf 22},
            178 (1975) [JETP Lett., {\bf 22}, 82 (1975)].
\bibitem{2} D.~K.~Campbell, J.~F.~Schonfeld, C.~A.~Wingate,
            Physica D, {\bf 9}, 1 (1983).
\bibitem{3} T.~I.~Belova, A.~E.~Kudryavtsev, Usp. Fiz. Nauk, {\bf 167} (4),
            377 (1997) [Physics -- Uspekhi, {\bf 40} (4), 359 (1997)].
\bibitem{4} D.~K.~Campbell, M.~Peyrard, P.~Sodano, Physica D, {\bf 19},
            165 (1986).
\end{thebibliography}
\end{document}